      [1999/12/01 v1.4c Il Nuovo Cimento]
\documentclass{cimento}

\usepackage{graphicx}  
\title{Nucleon electromagnetic structure: past, present, and future.}
\author{Egle Tomasi-Gustafsson\from{ins:x}}
\instlist{\inst{ins:x} DAPNIA-SPhN, CEA Saclay, F-91191 Gif-sur-Yvette}
\PACSes{\PACSit{25.30.Bf}{ Elastic electron scattering}{13.40.Gp}{ Electromagnetic form factors}}
\begin{document}

\maketitle

\begin{abstract}
We present the experimental status of electromagnetic hadron form factors. New and surprising results, based on polarization measurements, have been recently obtained for the electric proton and neutron form factors. In particular, the electric and magnetic distributions inside the proton appear not to be the same, in disagreement with results extracted from the unpolarized cross section, using the Rosenbluth separation. The new findings have given rise to a large number of papers and different speculations, as they question directly the models of nucleon structure and the reaction mechanism itself (based on $1\gamma$-exchange), with a possible revision of the calculation of radiative corrections, two-photon contribution etc. New data in time-like region are also available, through annihilation reactions. A large interest in this field arises, due also to the possibility of new measurements in polarized electron nucleon elastic scattering  at JLab, and also in the time-like region, at Frascati and at the future FAIR international facility.
\end{abstract}
\section{Introduction}

Form factors (FFs) characterize the internal structure of composite particles.  
They constitute a convenient playground for theory and experiment, because, on one side, they are directly related to experimental observables as cross sections and polarization observables, and, on the other side, they enter in the expression of the nucleon electromagnetic current, calculable by the models which describe the nucleon structure.

In a P and T invariant theory, a particle with spin $S$ is characterized by $2S+1$ electromagnetic form factors. The nucleon has two FFs, called electric ($G_{EN}$) and magnetic ($G_{MN}$), which, a priori, are different. Proton and neutron FFs are also different. 

Electromagnetic probes are traditionally preferred to the hadronic beams, and elastic electron hadron scattering contains all information on the nucleon ground state. The reaction mechanism is assumed to be one photon exchange, the electromagnetic interaction is exactly calculable in QED,  and one can safely extract the information on the hadronic vertex. However, one has to introduce radiative corrections, which become very large as the momentum transfer squared, $Q^2$, increases. Radiative corrections were firstly calculated by Schwinger \cite{Shwinger} and are important for the discussion of the experimental determination of the differential cross section. They are also calculable in QED. 

The four momentum transfer squared transmitted to the virtual photon, $q^2=-Q^2=-4EE'sin^2(\theta_e/2)$ (where  $E$ and 
$E^{\prime}$ are the energies of the incident and scattered electron, and $\theta_e$ is the electron scattering angle in Lab system), is negative in elastic $eh$ scattering, i.e., the space component is larger than the time component. The accessible kinematical region is called space-like (SL) region. Annihilation reactions, such as $N+\overline{N}\leftrightarrow \ell^+ +\ell^-$, $\ell=\mu$ or $e$, allow to scan the time-like (TL) region, $ q^2> 4m^2$, where $m$ is the nucleon mass. The relevant variable in TL region is the square of the total energy, $s=q^2$.

Form factors are analytical functions of $q^2$, 
being real functions in the SL  region (due to the hermiticity of the 
electromagnetic Hamiltonian) and complex functions in the 
TL region. The Phr\`agmen-Lindel\"of 
theorem \cite{Ti39} gives a rigorous prescription for the asymptotic behavior of analytical functions: 
$\lim_{q^2\to -\infty} F^{(SL)}(q^2) =\lim_{q^2\to \infty} 
F^{(TL)}(q^2)$.
This means that, asymptotically, FFs, have the following constraints: 
1) the time-like phase vanishes  and 2) the real part of FFs, 
${\cal R}e  F^{(TL)}(q^2)$, coincides with the 
corresponding value, $F^{(SL)}(q^2)$.

These asymptotic properties based on analiticity, however, are different from the asymptotics properties of FFs, predicted in QCD, which derive from scaling rules and helicity conservation. Therefore, the study of FFs at large $Q^2$ represents a unique tool for the understanding of these properties of the nucleon dynamics.

\section{The past}

The importance of FFs was recognized since the first measurements  of R. Hofstadter \cite{Ho62}, Nobel laureate in 1961, to whom  G. Liljestrand, member of the Royal Academy of Sciences, addressed with these words: {\it " The myth of the indivisibility of the atom, implied in its very name, was shattered in the beginning of this century, and a completely new and fascinating world of the utmost importance became revealed. You have been able to obtain further significant information of the intimate structure of this intriguing world by disclosing the distribution of electric charges and magnetic forces within the atomic nucleus, and the particles of which it is composed." }

An elegant formalism, in QED, allows to relate measured quantities, as cross section and polarization observables to FFs. The expressions which relate the moduli of FFs to the unpolarized differential cross section were developed by Rosenbluth \cite{Ro50} for the scattering channel,  $e+N\to e+N$, and, for the annihilation channel, by Zichichi, Berman, Cabibbo and  Gatto \cite{Zi62}.  The importance of polarization phenomena, and the related formalism, were firstly suggested by the 'Kharkov school' of Akhiezer and Rekalo \cite{Re68}, and, in TL region, by Bilenkyi, Giunti and Wataghin \cite{Bi93} and  Dubnickova, Rekalo and Dubnicka \cite{Du96}.

The comparison of the observables with the theoretical models of hadron structure is straightforward, as, in framework of the mechanism of one-photon approximation, FFs enter directly in the expression of the hadronic electromagnetic current. The matrix element of the process $e+N\to e+N$ is written as:
\begin{equation}
{\cal  M}_1 =\displaystyle\frac {e^2}{Q^2}\overline{u}(k_2)\gamma_{\mu}u(k_1) \overline{u}(p_2)\left [F_{1N}(Q^2)\gamma_{\mu}-
\displaystyle\frac{\sigma_{\mu\nu}q_{\nu}}{2m}F_{2N}(Q^2)\right] u(p_1),
\label{eq:mat}
\end{equation}
where $k_1$ $(p_1)$ and $k_2$ $(p_2)$ are the four-momenta of the initial and final electron (nucleon),  $q=k_1-k_2$, $Q^2=-q^2>0$. $F_{1N}$ and $F_{2N}$ are the Dirac and Pauli nucleon electromagnetic form factors. The same FFs enter also in the description of elastic scattering of positrons by nucleons. From Eq. (\ref{eq:mat}) one can find the following expression for the differential cross section in the laboratory  system:
\begin{equation}
\displaystyle\frac{d\sigma}{d\Omega}_e=\sigma_0\left [ G_{MN}^2(Q^2)+
\displaystyle\frac{\epsilon}{\tau}G_{EN}^2(Q^2)\right ],~\tau=Q^2/(4m^2),
\label{eq:csst}
\end{equation}
where $\sigma_0$ is a kinematical factor, which contains the Mott cross section, for the scattering of unpolarized electrons by a point charge particle (with spin 1/2), 
$\epsilon$ is the second independent kinematical variable, which, together with $Q^2$,  fully determines the kinematics of  elastic $eN$-scattering and can be written, in the zero electron mass limit, as:
\begin{equation}
\epsilon=
\left [1+2(1+\tau)\tan^2 \displaystyle\frac{\theta_e}{2}\right ]^{-1},~0\le \epsilon\le 1.
\label{eq:csst1}
\end{equation}
The Sachs FFs $G_{MN}$ and $G_{EN}$ are related to the Dirac and Pauli FFs by: $G_{MN}=F_{1N}+F_{2N}$, and $G_{EN}=F_{1N}-\tau F_{2N}$.
From Eq. (\ref{eq:csst}) it appears that the cross section depends linearly on $\epsilon$, and measurements at fixed $Q^2$, at different angles, allow a straightforward extraction of  the electric and magnetic FFs.  However, as $Q^2$ increases, the electric contribution becomes small, compared to the magnetic part, which is weighted by the factor $\tau$. Its precise determination becomes, therefore, more difficult.

Taking a longitudinally polarized electron beam, and measuring the polarization of the outgoing proton, in $\vec e+p\to e+\vec p$ (or, alternatively, using a polarized target, with longitudinally polarized beam), the polarization induces an interference term, in the cross section, related to the product of $G_{MN}$ and $G_{EN}$. It has been shown that the  transverse ($P_T$) and the longitudinal ($P_L$)  polarization of the proton in the reaction plane (the third component, normal to the scattering plane vanishes, due to 
T-invariance) are proportional to $G_{Ep}G_{Mp}$ and $G_{Mp}^2$, respectively, so that the simultaneous measurements of these two polarization components gives directly the ratio of the form factors:
\begin{equation}
\displaystyle\frac{G_{Ep}}{G_{Mp}}=-\displaystyle\frac{P_T}{P_L}
\displaystyle\frac{\left (E+E^{\prime} 
\right )}{2m}\tan \frac{\theta _e}{2}.
\label{ffratio}
\end{equation} 
The experimental realization of this method requires a polarized beam with high intensity, as the proton polarization has to be measured through a secondary scattering on a carbon or polyethylene target. This is the reason for which it has become possible only in the recent years.

\section{The present}

\subsection{Space-like region}

In SL region, the measurements based on the Rosenbluth method show that the behavior of the magnetic proton and neutron FFs as a function of $Q^2$, follows approximately a dipole law:
\begin{equation}
G_{MN}(Q^2)/\mu_N=G_d,~\mbox{with}~
G_d=\left [1+{Q^2}/{ m_d^2 }\right ]^{-2},~m_d^2=0.71~\mbox{GeV}^2,
\label{eq:dipole}
\end{equation}
where $\mu_N$ is the nucleon magnetic moment in units of the Bohr magneton, respectively $\mu_p=2.79$ for proton  and $\mu_n=-1.913$ for neutron. 

Such behavior is consistent with the scaling laws predicted by QCD \cite{Le80}, but also with a non relativistic picture of an exponential distribution of the magnetization inside the nucleon.

Concerning the electric proton FF, when comparing the data derived from the Rosenbluth separation and from the polarization transfer method, it turns out that a discrepancy appears between the $Q^2$-dependences of the FFs ratio $R$.

From the Rosenbluth separation data one finds the scaling relation:
$R= \mu_p G_{Ep}/G_{Mp}$ $\simeq 1$, whereas the following parametrization describes the polarization data \cite{Jo00,Ga02}:
\begin{equation}
R=\mu_p G_{Ep}/G_{Mp}=1-0.13(Q^2~[\mbox{GeV}^2]-0.04)
\label{eq:brash}
\end{equation}
which implies that the ratio monotonically decreases and deviates from unity, as $Q^2$ increasing, reaching a value of $R\simeq$ 0.3 at $Q^2= 5.5 $ GeV$^2$.

A careful experimental and theoretical analysis of this problem is necessary. The important point is the calculation of radiative corrections to the differential cross section and to polarization observables in elastic $eN$-scattering. If these corrections are large (in absolute value) for the differential cross section \cite{Mo69}, in particular for high resolution experiments, a simplified estimation of radiative corrections to polarization phenomena \cite{Ma00} shows that radiative corrections are small for the ratio $P_T/P_L$. The possibility of a contribution of two photon exchange, suggested long ago \cite{Gu73}, is actually investigated, and gives rise to experimental \cite{04019,Novosibirsk} and theoretical efforts \cite{Twof,Re04}.

Experiments, based on the Rosenbluth method have been performed in several laboratories, (SLAC, Bonn, DESY..), especially at low $Q^2$. A recent analysis and exhaustive references on proton FFs experiments can be found in \cite{Ar03}. 

Experiments, based on the polarization transfer method, have been performed at MAMI, NIKHEF, JLab... up to $Q^2=5.6$ GeV$^2$ \cite{Jo00,Ga02}.  

The neutron electric FF is small and  the polarization transfer method appears very useful, even at low momentum transfer. Recent measurements, using a polarized target ($^3\!He$ or  deuterium), or measuring the polarization of the ougoing nucleon (\cite{Glazier} and refs. herein), have been done up to $Q^2$=1.8 GeV$^2$ and show that $G_{En}$ is definitely different from zero.

The world data are shown in Fig. 1. The proton electric FF data obtained from unpolarized measurements are shown in  Fig. 1a, (stars), whereas the polarization data are shown as solid squares. The discrepancy among the data issued from the two methods appears clearly.

Different models exist for the description of the nucleon structure. We show, here, as an example, predictions from pQCD, from VDM inspired models \cite{Ia73,Lo02} and from an empirical fit \cite{Bo94}. For model \cite{Ia73} (dotted line) and for model \cite{Bo94} (dashed line) the parameters are the same as in the original work, but for model \cite{Lo02}  (solid line), a better representation could be obtained after a global fit on all data in SL and TL regions. The pQCD prediction, which follow the dipole law (except for $G_{En}$) is shown as dot-dashed line.

\begin{figure}[pht]
\begin{center}
\includegraphics[width=16cm]{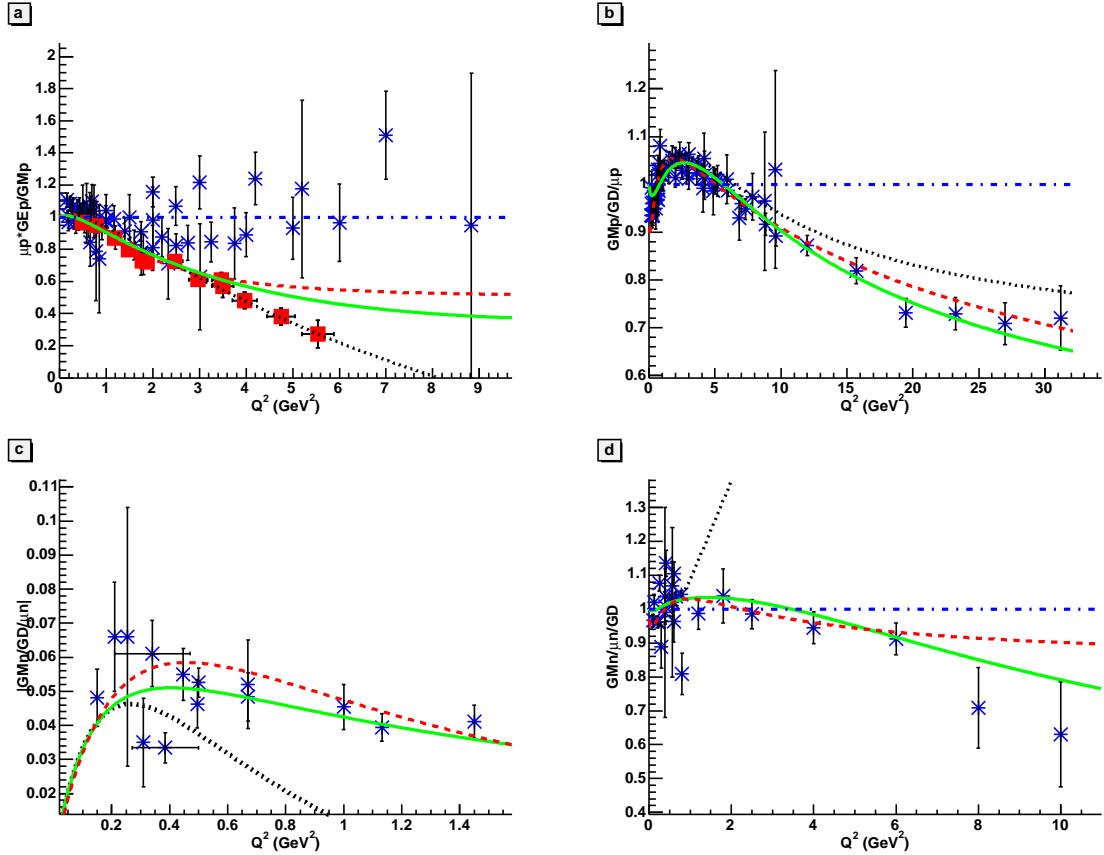}
\caption{\label{fig:fig1} Nucleon Form Factors in Space-Like region: (a) proton electric FF, scaled by $\mu_p G_{Mp}$ (b) proton magnetic FF, (c) neutron electric FF, (d) neutron magnetic FF, scaled by $\mu_n G_{Mn}$. The predictions of the models are drawn: pQCD (dot-dashed line), model from Ref. \cite{Ia73} (dashed line), model from Ref. \cite{Lo02} (solid line), model from Ref. \cite{Bo94} (dotted line). }
\end{center}
\end{figure}

\subsection{Time-like region}

In TL region, the measurement of the differential 
cross section for the processes $\overline{p}+p\leftrightarrow \ell^+ +\ell^-$ at a fixed value of the square of the total energy $s$ and for two different angles of the scattered particle, $\theta$, allows the separation of the two FFs, $|G_M|^2$ and $|G_E|^2$, and it is equivalent to the Rosenbluth separation for 
the elastic $ep$-scattering. This procedure is simpler in TL region, as it 
requires to change only one kinematical variable, $\cos\theta$, whereas, in SL region, 
it is 
necessary to change simultaneously two kinematical variables: the energy of the 
initial electron and the electron scattering angle, fixing the momentum transfer squared, $q^2$. However, the Rosenbluth 
separation  of the $|G_E|^2$ and $|G_M|^2$ contributions, has not been realized yet due to the limited statistics which is possible to achieve.

In order to determine the form factors \cite{An03}, the differential cross section has to be integrated over a wide angular 
range. One typically assumes that the $G_E$-contribution plays a minor role in the cross 
section at large $q^2$ and the 
experimental results are usually given 
in terms of $|G_M|$, under the hypothesis that $G_E=0$ or $G_E=G_M$. The first hypothesis is arbitrary. The second hypothesis is strictly 
valid at threshold only, i.e., for $\tau=1$, but there is no 
theoretical argument which justifies its validity at any other momentum 
transfer. 

\begin{figure}[pht]
\begin{center}
\includegraphics[width=16cm]{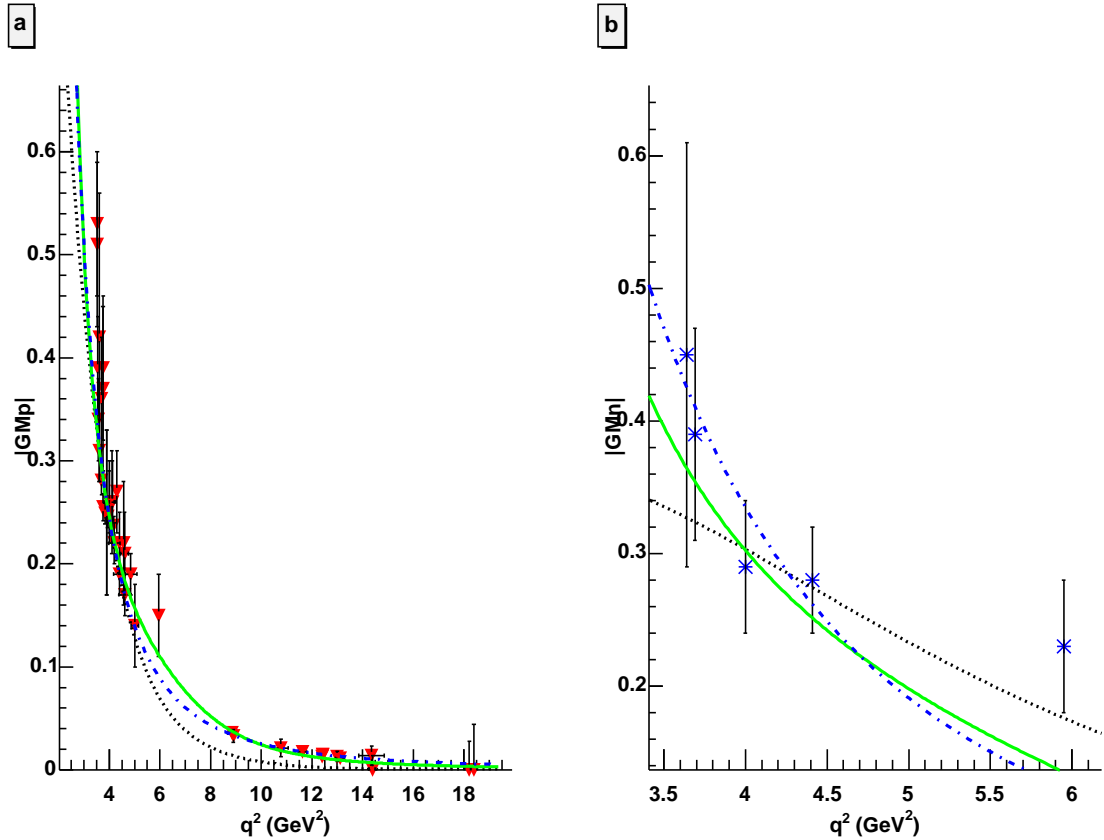}
\caption{\label{fig:fig2} Form Factors in Time-Like region. Different figures and curves are described in the text.}
\end{center}
\end{figure}
The $|G_M|^2$ values depend, in principle, on the kinematics where the 
measurement was performed and on the angular 
range of integration, however it turns out that these two assumptions for $G_E$ 
lead to values for $|G_M|$ which differ by 20\% at most.

Few data exist, especially at large $s$,  in the TL region, for proton and even less for neutron . FFs in the TL region are larger 
than the 
corresponding SL data. This has been 
considered 
as a proof of the non applicability of the Phr\`agmen-Lindel\"of theorem, (up to $s$=18 GeV$^2$, at least) or as an evidence that the asymptotic regime is not reached \cite{Bi93}. 

The data in TL region, in the hypothesis that $G_E=G_M$ are shown in Fig 2a for the proton and in Fig 2b for the neutron, together with the predictions of models. Proton FFs have been measured in $p+\overline p\to e^+ + e^-$ at CERN and Fermilab, in $e^+e^-\to p+\overline p$ at Orsay and Frascati \cite{An03}. Neutron TL FFs have been measured at Frascati \cite{An98}.

The extension to TL region of the VDM inspired models \cite{Ia73,Lo02} is based on the following relations, which are necessary for the analytical continuation between SL and TL regions:
\begin{equation}
Q^2=-q^2=q^2e^{-i\pi}~\Longrightarrow~\left\{\begin{array}{c}
\ln(Q^2)=ln(q^2)-i\pi\\
\sqrt{Q^2}=e^{\frac{-i\pi}{2}}\sqrt{q^2}\\
\end{array} \right.
\end{equation}

The pQCD prediction (\ref{eq:dipole}) can be extended in TL region as \cite{Le80}:
\begin{equation}
|G_M|=\frac{A}{s^2\ln^2(s/\Lambda^2)},
\label{eq:tlpqcd}
\end{equation}
where $\Lambda=0.3$ GeV is the QCD scale parameter and $A$ is a free parameter.
This simple parametrization is taken to be the same for proton and neutron. The best fit is obtained with a parameter $A(p)$= 56.3 for proton and 
$A(n)$= 77.15 for neutron. This reflects the fact that in TL region, neutron FFs are larger than for proton, although the errors are also larger. It has been suggested that an $N\overline{N}$ bound state just below the $N\overline{N}$ threshold could be responsible for this difference \cite{Ga96}.

\section{The future}

In SL region, if the trend suggested by the recent data based on the polarization metod is confirmed, an extension of the measurements planned at JLab up to 9 GeV$^2$ will show evidence for a zero crossing of the electric FF, which could even, eventually, become negative \cite{04108}. 

Concerning the neutron, the extension of large $Q^2$ of the measurement on the electric FF will confirm or infirm the fact that this quantity is larger than previously assumed. 

These 'surprises' on the nucleon FFs affect the description of the light nuclei structure, such as deuteron, and demand for a reevaluation of the role of the different ingredients usually taken to calculate their structure, such as meson exchange current, relativistic corrections ..\cite{ETG01}.

In TL region, 
the angular dependence of the differential cross section for $\overline{p}+p\to \ell^+ +\ell^-$ as a 
function of the angular asymmetry ${\cal R}$ is:
\begin{equation}
\displaystyle\frac{d\sigma}{d(\cos\theta)}=
\sigma_0\left [ 1+{\cal R} \cos^2\theta \right ],~
{\cal R}=\displaystyle\frac{\tau|G_M|^2-|G_E|^2}{\tau|G_M|^2+|G_E|^2},
\label{eq:eq3}
\end{equation}
where $\sigma_0$ is the value of the differential cross section at 
$\theta=\pi/2$. 

The angular dependence of the cross section, Eq. (\ref{eq:eq3}), results 
directly from the assumption of one-photon exchange, where the spin of the 
photon 
is equal 1 and the electromagnetic hadron interaction satisfies the 
$C-$invariance. 
Therefore, the measurement of the differential 
cross section at three angles (or more) would also allow to test the presence of 
$2\gamma$ exchange \cite{Re99}. 

Polarization phenomena will be especially interesting in $\overline{p}+p\to \ell^+ +\ell^-$. For example, the transverse polarization, $P_y$, of the proton target (or the transverse polarization of the antiproton beam) results in nonzero analyzing power \cite{Zi62,Bi93}:
$$
\displaystyle\frac{d\sigma}{d\Omega}(P_y)=
\left ( \displaystyle\frac{d\sigma}{d\Omega} \right )_0 \left [1+{\cal A}P_y
\right ],
$$
$${\cal A}=\displaystyle\frac{\sin 2\theta Im G_E^*G_M}{D\sqrt{\tau}},
~D=|G_M|^2(1+\cos^2\theta)+\displaystyle\frac{1}{\tau}|G_E|^2\sin^2\theta.$$
This analyzing power characterizes the T-odd correlation $\vec P\cdot\vec k\times\vec p$, where $\vec k(\vec p)$ is the three momentum of the $\overline{p}$ beam (produced lepton). 

\begin{figure}[pht]
\begin{center}
\includegraphics[width=14cm]{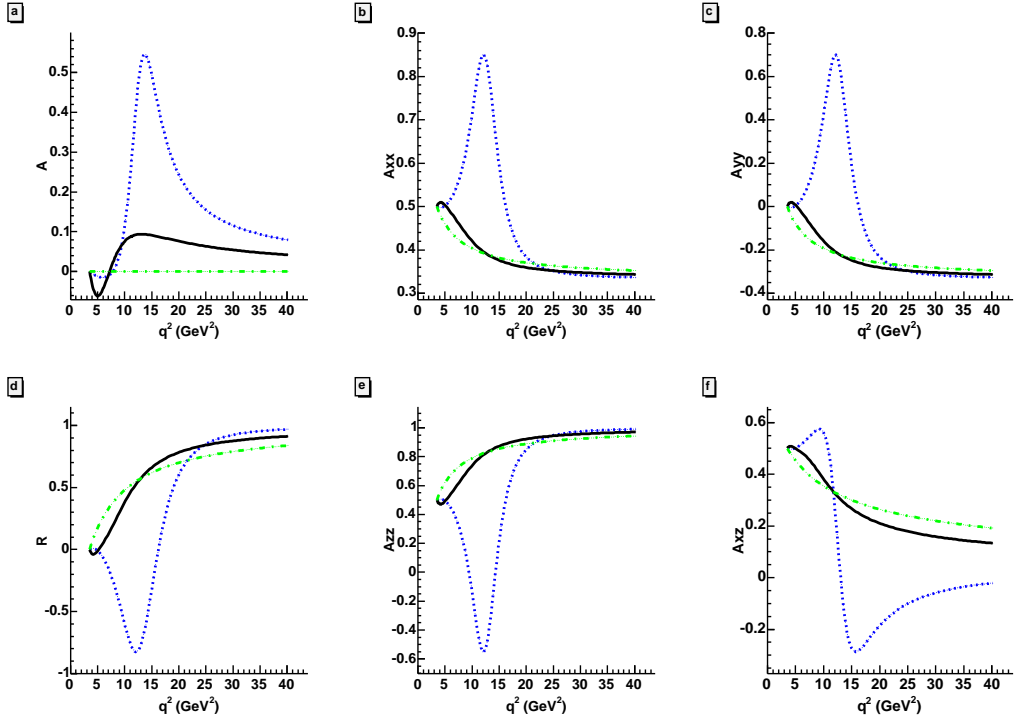}
\caption{\label{fig:fig3} Polarization observables. Different figures and curves are described in the text.}
\end{center}
\end{figure}

The same information can be obtained from the final polarization in $\ell^++\ell^- \to \vec p+\overline{p}$, but in this case one has to deal with the problem of hadron polarimetry, in conditions of very small cross sections.
When both colliding particles are polarized, one acccess to nine possible double spin observables, $A_{ab}$ (where $a$ and $b=x,y,z$ refer to the $a(b)$ component of the target(projectile) polarization) 
four of which vanish: $A_{xy}=A_{yx}=A_{zy}=A_{yz}=0$. The nonzero components are:
\begin{eqnarray*}
\displaystyle\frac{d\sigma}{d\Omega}A_{xx}&=& 
\sin^2\theta\left (|G_M|^2 +\displaystyle\frac{1}{\tau}|G_E|^2\right ){\cal N},\\
\displaystyle\frac{d\sigma}{d\Omega}A_{yy}&=& 
-\sin^2\theta\left (|G_M|^2 -\displaystyle\frac{1}{\tau}|G_E|^2\right ){\cal N},\\
\displaystyle\frac{d\sigma}{d\Omega}A_{zz}&=& 
\left [(1+\cos^2\theta)|G_M|^2-
\displaystyle\frac{1}{\tau}\sin^2\theta |G_E|^2\right ]{\cal N},\\
\displaystyle\frac{d\sigma}{d\Omega}A_{xz}&=&\displaystyle\frac{d\sigma}{d\Omega}A_{zx}=\displaystyle\frac{1}{\sqrt{\tau}}\sin 2\theta Re G_E G_M^* {\cal N}.
\end{eqnarray*}
where 
${\cal N}=\displaystyle\frac{\alpha^2}{4\sqrt{t(t-4m^2)}}$, $\alpha=e^2/(4\pi)\simeq 1/137 $ is a kinematical factor.
The predictions for TL observables are shown in Fig. \ref{fig:fig3} for the models described above: the cross section asymmetry, the single spin asymmetry, $A$, (Fig. \ref{fig:fig3}a), the angular asymmetry, ${\cal R}$, (Fig. \ref{fig:fig3}d),  and the double spin polarizations $A_{xx}$ (Fig. \ref{fig:fig3}b), $A_{yy}$ (Fig. \ref{fig:fig3}c),$A_{zz}$ (Fig. \ref{fig:fig3}e) and $A_{xz}$ (Fig. \ref{fig:fig3}f). For all these  observables a large difference appears according to the different models, which qualitatively describe the available SL and TL data. In particular, even the sign can be opposite for VDM inspired models and pQCD. The model \cite{Lo02}, is somehow intermediate between the two representations, as it includes the asymptotic predictions of QCD, at the expenses of a larger number of parameters. It is important to note that the $\tau$-dependence of ${\cal A}$ is very sensitive to existing models of the nucleon FFs, which reproduce equally well the data in SL region \cite{Br03}.

\section{Conclusions}

The field and the formalism of nucleon electromagnetic FFs, developped a few decades ago, is still source of very interesting developments. 

A unique and satisfactory interpretation of the four nucleon FFs (electric and magnetic, for neutron and proton) in TL and SL momentum transfer region 
has not yet been reached.

In SL region the precise determination of proton and neutron FFs at large momentum transfer will further strongly constrain the models on light nuclei structure.

Interesting problems will be adressed and solved by future measurements in TL region:
\begin{itemize}
\item
the separation of the electric and magnetic FFs, through the angular distribution of the produced leptons: an interesting observable is the measurement of the asymmetry ${\cal R}$ (from the angular dependence 
of the 
differential cross section for $\overline{p}+p\leftrightarrow \ell^+ +\ell^-$) which is 
sensitive to the relative value of $G_M$ and $G_E$ and does not require polarization observables.
\item
the presence of a large relative phase of magnetic and electric proton FFs 
in the TL region, if experimentally proved at relatively large momentum transfer using polarized target (or beam), will be a strong  
indication that these FFs have a different behavior.

\item the study of the processes $\overline{p}+p\to \pi^0+ \ell^+ +\ell^-$ and $\overline{p}+p\to \pi^++\pi^-+\ell^+ +\ell^-$, will allow to measure proton FFs in the TL region, for $s\le 4m^2$, where the vector meson contribution plays an important role.
\end{itemize}
It will be possible, in recent future, to achieve a new level of precision and also to explore kinematical regions where data are totally absent. This program is especially interesting with respect to the important problem of  the transition to the asymptotic region,  predicted by QCD, which actually gives rise to many discussions and speculations. FFs should sign unambiguosly the 'transition region', where quarks degrees of freedom should be taken explicitely into accounts and the description of the nucleon, in terms of meson and effective degrees of freedom, no longer holds.

\acknowledgments

This work would not have been done without many useful discussions and suggestions from Prof. Michail P. Rekalo.

Thanks are due to Florent Lacroix and Charles Duterte for essential help in the numerical calculations and to Dr. Gennadiy I. Gakh for a careful reading of the manuscript.


\begin{thebibliography}{0}


\bibitem{Shwinger}
\BY{Schwinger~ J.~S.}
\IN{Phys. Rev.}{76}{1949}{790}.

\bibitem{Ti39} 
\BY{Titchmarsh~E.~C.} , \TITLE{ Theory of functions}, edited by  Oxford University 
Press, London, 1939.

\bibitem{Ho62} 
\BY{Hofstadter~R., Bumiller~F \atque Yearian~M.} 
\IN{Rev. Mod. Phys.}{30}{1958}{482}.

\bibitem{Ro50} 
\BY{Rosenbluth~M.~N.}
\IN{Phys. Rev.}{79}{1950}{615}.
\bibitem{Zi62}  
\BY{Zichichi~A., Berman~S.~M., Cabibbo~N.  \atque  Gatto~R.} 
\IN{Nuovo Cimento}
{ XXIV}{1962}{170}. 
\bibitem{Re68} 
\BY{Akhiezer~A.  \atque  Rekalo~M.~P.~} 
\IN{Dokl. Akad. Nauk USSR}{180}{1968}{1081}; 
\IN{Sov. J. Part. Nucl.}{4}{277}{1974}.
\bibitem{Bi93} Bilenky~S.~M., Giunti~C. \atque  Wataghin~V.
\IN{Z. Phys.C}{59}{1993}{475}.
\bibitem{Du96} 
\BY{Dubnickova~A.~Z.,~Rekalo~M.~P.  \atque Dubnicka~S.} 
\IN{Z. Phys. C}{70}{1996}{473}.
\bibitem{Le80}
\BY{Lepage~G.~P. \atque  Brodsky~S.~J.}
\IN{Phys. Rev. D}{22}{1980}{2157};
\IN{Phys. Rev. Lett.}{43}{1979}{545}; 
\IN{[Erratum-ibid.]}{43}{1979}{1625}.

\bibitem{Jo00}
\BY{Jones~M.~K. et al.}  [Jefferson Lab Hall A Collaboration],
\IN{Phys. Rev. Lett.}{84}{2000}{1398}.
\bibitem{Ga02}
\BY{Gayou~O et al.}  [Jefferson Lab Hall A Collaboration],
\IN{Phys. Rev. Lett.}{88}{2002}{092301}.
\bibitem{Mo69} 
\BY{Mo~L.~W. \atque Tsai ~Y.~S.}
\IN{ Rev. Mod. Phys.}{41}{1969}{205}.
\bibitem{Ma00} 
\BY{L.~C.~Maximon \atque  J.~A.~Tjon}
\IN{Phys. Rev. C}{62}{2000}{054320};\\
\BY{Afanasev~A., Akushevich~I. \atque  Merenkov~N.}
\IN{Phys. Rev. D}{64}{2001}{113009};
\BY{Afanasev~A.~V.,Akushevich~ I., Ilyichev~A. \atque  Merenkov~N.~P.}
\IN{Phys. Lett. B}{514}{2001}{269}; 
\bibitem{Gu73}  
\BY{ Gunion~J. \atque  Stodolsky~L.}  
\IN{Phys. Rev. Lett.}{30}{1973}{345};
\BY{Franco~V. }  
\IN{Phys. Rev. D }{8}{1973}{826}; 
\BY{Boitsov~V.~N. , Kondratyuk~L.~A.  \atque  Kopeliovich~V.~B.}
\IN{Sov. J. Nucl. Phys.}{16}{1973}{237};
\BY{Lev~F.~M.}
\IN{Sov.~J.~Nucl.~Phys.}{21}{1973}{45}.

\bibitem{04019}
\BY{Gilman~R., Pentchev~L., Perdrisat~C.~F., Suleiman ~R.} 
{\it JLab Proposal 04-019}, 2004.

\bibitem{Novosibirsk}
\BY{Arrington~J. et al.}
{\it arXiv:nucl-ex/0408020}.

\bibitem{Twof}
\BY{Blunden~P.~G., Melnitchouk~W.\atque  Tjon~J.~A.}
\IN{Phys. Rev. Lett.}{91}{2003}{142304};
\BY{Guichon~P.~A.~M. \atque  Vanderhaeghen~M.~}
\IN{Phys. Rev. Lett.}{ 91}{2003}{142303};
\BY{Chen~Y.~C., Afanasev~A., Brodsky~S.~J., Carlson~C.~E. \atque  Vanderhaeghen~M.}
\IN{Phys. Rev. Lett.}{93}{2004}{122301}.
\bibitem{Re04}
\BY{Rekalo~M.~P. \atque Tomasi-Gustafsson~E.}
\IN{Eur. Phys. J. A}{22}{2004}{331};
\IN{Nucl. Phys. A}{740}{2004}{271};
\SAME{742}{2004}{322}.


\bibitem{Ar03}
\BY{Arrington~J.}
\IN{Phys. Rev. C}{68}{2003}{034325}.

\bibitem{Glazier}
\BY{Glazier~D.~I. et al.}
{\it arXiv:nucl-ex/0410026}.


\bibitem{Ia73}
\BY{Iachello~F., Jackson~A.~D. \atque Lande~A}.,
\IN{Phys. Lett. B}{43}{1973}{191}; 
\BY{F.~Iachello \atque  Q.~Wan}
\IN{Phys. Rev. C}{69}{2004}{055204}.

\bibitem{Lo02}
\BY{Lomon~E.~L.}
\IN{Phys. Rev. C}{66}{2002}{045501}.
\bibitem{Bo94}
\BY{Bosted~P.~E.}
\IN{Phys. Rev. C}{51}{1995}{409}.
\bibitem{An03} 
\BY{Andreotti~M.  et al.}  
\IN{Phys. Lett. B}{559}{2003}{20} and refs herein.


\bibitem{An98}
\BY{Antonelli~A. et al.}
\IN{Nucl. Phys. B}{517}{1998}{3}.

\bibitem{Ga96}
\BY{Gauzzi~P.}
\IN{Phys. Atom. Nucl.}{59}{1996}{1382}
[\IN{Yad. Fiz.}{59}{1996}{1441}].
\bibitem{04108} 
\BY{Brash~E., Jones~M., Perdrisat~C.~F., Punjabi~V.} 
{\it JLab Proposal 04-108}, 2004.
\bibitem{ETG01} 
\BY{Tomasi-Gustafsson~E. \atque  Rekalo~M.~P.} 
\IN{Phys. Lett. B}{504}{2001}{291} and  refs herein.
\bibitem{Re99} 
\BY{Rekalo~M.~P., Tomasi-Gustafsson~E.  \atque Prout~D.} 
\IN{Phys. Rev. C}{60}{1999}{042202}.
\bibitem{Br03} 
\BY{Brodsky~S.~J., Carlson~C.~E., Hiller~J.~R. \atque  Hwang~D.~S.}
\IN{Phys. Rev. D}{69}{2004}{ 054022}.

\end{thebibliography}
\end{document}